\documentclass[11pt,fleqn]{article}
\usepackage{epstopdf} 
\usepackage{graphicx}
\usepackage[a4paper,left=2cm,right=2cm]{geometry}
\usepackage{amsmath}
\usepackage{float}
\usepackage[latin1]{inputenc}
\usepackage{amsmath}
\usepackage{amsfonts}
\usepackage{amssymb}
\usepackage{float}
\usepackage[toc,page]{appendix}
\floatstyle{boxed} 
\restylefloat{figure}

\newcommand{\be}{\begin{equation}}
\newcommand{\ee}{\end{equation}}
\newcommand{\bea}{\begin{eqnarray}}
\newcommand{\eea}{\end{eqnarray}}

\begin{document}
\title{Exotic Leptonic solutions to observed anomalies in lepton universality observables and more.\footnote{based on a Talk given at 23rd DAE-BRNS High Energy Physics Symposium 2018, IIT Chennai, 10-14 Dec. 2018.}}
\author{Lobsang Dhargyal \\\\\ Harish-Chandra Research Institute, HBNI, Chhatnag Road, Jhusi Allahabad 211 019 India.}

\maketitle
\begin{abstract}
In this talk I will present the work that we did in \cite{1}\cite{2}\cite{3}\cite{4}\cite{5} related to observed lepton universality violation by Babar, Belle and LHCb in R($D^{(*)}$) and $R_{K^{(*)}}$ as well as the reported deviation in muon (g-2) by BNL. We had shown that all these anomalies as well as Baryon-genesis, Dark-matter and small neutrino masses could be explained by introducing new exotic scalars, leptons and scalar-leptoquarks only. It turn out that some of these models have very peculiar signatures such as prediction of existence of heavy stable charged particle \cite{1}\cite{2}, vector like fourth generation leptons \cite{3} or even scalar Baryonic DM candidates etc. Some of these models turn out to have very unique collider signatures such as $ee/pp \rightarrow \mu\mu(\tau\tau)\ +\ missing\ energy\ (ME)$, see \cite{1}\cite{2}\cite{4}. This is interesting in the sense that such peculiar signatures of these new particles can be searched in the upcoming HL-LHC or with even better chance of observing these signatures are in the upcoming precision machines such as ILC, CEPC etc.

\end{abstract}

In this talk I will present a short summary of interesting consequences of introducing new exotic leptons and scalars (LQ) to resolve the reported anomalies by Babar, Belle, LHCb, BNL in R($D^{(*)}$), $R_{K^{(*)}}$, muon (g-2). In this brief talk the emphasis is laid on the peculiar features (and interesting side observations) of the models that we proposed to resolve the mentioned anomalies, for more details we refer the readers to the sources \cite{1}\cite{2}\cite{3}\cite{4}\cite{5}. In the following we present the particle content, achievements and peculiar features and observations about the respective models.\\
\\
(1) : In \cite{1} we extended the inert-doublet 2HDM (IDM) by introducing three exotic leptons ($F_{iR,L}$ for i = 1, 2, 3) which is singlet under the SM $SU(2)_{L}$ and vector like under the SM $U(1)_{Y}$ beside a new $U(1)_{F}$ to which only the right handed new exotic leptons are charged (all exotic leptons are odd under the $Z_{2}$ to avoid very stringent tree level constrains). This model can explain the reported anomaly in muon (g-2) as well as why the anomaly is observed only in the muon sector and not in the electron sector (due to a peculiar solution choice of the $\gamma_{5}$ anomaly cancellation in \cite{1}) besides small neutrino masses and Baryogenesis via leptogenesis (same as in the Scoto-genic model). But in \cite{2} we realized that if let the left handed of the exotic fermions charged under the new $U(1)_{F}$ instead of the right handed one as in \cite{1}, then beside the model explaining the anomalies explained by model in \cite{1}, it is also able to incorporate the observed deviations in $R_{K^{(*)}}$ as well (via box loop diagrams). Due to a peculiar choice of $\gamma_{5}$ anomaly free conditions in \cite{1}\cite{2}, which besides explaining why the anomlay is in the muon sector and not in the electron sector as mentioned before, it also predict one stable (long lived\footnote{unless there also exist doubly charged scalar, in which case it need not be stable}) charged exotic lepton which can be very heavy (in fact heavier than the unstable exotic lepton) which could turn up in LHC or HL-LHC data or even more prominently in ee colliders such as future ILC, CEPC etc. If (somehow) this stable lepton has same mass as electron then only weak interaction charged current will be able to differentiate between the SM electron and the exotic electron properly.\\
\\
(2) : Taking inspirations from \cite{1} and \cite{2}, in \cite{3} we built a NP model by introducing vector like exotic leptons and scalars (LQ) to explain the observed deviations in $R(D^{(*)})$ data at loop level. And in \cite{4} we have introduced pair of charged exotic lepton doublets ($L_{1L}$ and $L_{2R}$) and a pair of charged exotic lepton singlets ($E_{1R}$ and $E_{2L}$)\footnote{where doublets and singlets refers to SM gauge group $SU(2)_{L}$}, which is free of $\gamma_{5}$ anomaly, besides new LQ and scalars and shown that both the R($D^{(*)}$) and $R_{K^{(*)}}$ can be explained within the limits of present error estimates. One peculiar feature of the models in \cite{3} and \cite{4} is that to satisfy the very stringent constrains from the $K^{0}-\bar{K}^{0}$ and $B^{0}-\bar{B}^{0}$ mixings, we are force to restrict the CKM angles in $\pi \leq \theta_{12} \leq \frac{3\pi}{2}$ and $\frac{3\pi}{2} \leq \theta_{13},\theta_{23} \leq 2\pi$ and impose a constrains similar like GIM on the combinations of Yukawa couplings and CKM elements which lead to requirements of at least one of the Yukawa couplings must be complex which in fact predicts a small CP violation (in a particular choice of the Yukawa couplings here, but it could be made large too in other choices!) in $B^{0}_{s}-\bar{B}_{s}^{0}$ mixing due to new exotic leptons (box loop level), for details see \cite{3}\cite{4}.\\
\\
(3) : In \cite{5} we have proposed a NP model (two pair of $SU(2)_{L}$ singlet right handed leptons carrying opposite $U(1)_{Y}$ charges and their two left handed counterparts which are neutral singlet leptons) where in the regime where the exotic leptons masses are in the electroweak (EW) scale, the model will be able to explain the $R_{K^{(*)}}$ and muon (g-2) as well as small neutrino masses via
minimum-inverse seesaw scenario (MISS), for more details see \cite{5} and references there in. In this model when the exotic fermion masses are well above the EW scale then we can have stable scalar baryon (singlet under strong interaction) with charge -3 which could explain the primordial Li problem  by forming a hydrogen like atom with the $Li^{+3}$ nucleus with peculair absoprtion or emmision line of first excited state in the X-ray region at $E_{2}(Li) - E_{1}(Li) \approx 10.62$ MeV. Another peculiar but very interesting side observation about this model is that when the new fermion masses are well above the EW scale, then if we assign $U(1)_{Y}$ charges (unlike the way we did above) such that there are vector like under $U(1)_{Y}$, then it will not be able to explain the $R_{K^{(*)}}$ and muon (g-2) as well as MISS is not possible now, but in this case we can have the exotic leptons decay into stable scalar quarks. These stable scalar quarks (if we assign them a flavor $SU(3)_{F}$ similar/same like the SM quark u, d and s flavors) can have electromagnetically neutral (as well as $Q$ = +2 depending on the charges of the exotic fermions) stable scalar baryon and is expected to be suppressed under the strong interaction at the level of OZI rule or smaller (due to scalar baryon being singlet under both color and flavor and also due to heavy mass scale of scalar quarks and small size (due to lack of exlusion force)) which can be a DM candidate and also in these kind of models the origins of ordinary baryons and DM (scalar baryons here) could be linked.


\begin{thebibliography}{99}

\bibitem{1} Lobsang Dhargyal, \textsl{Eur.Phys.J. C78 (2018) no.2, 150 and references there in.}

\bibitem{2} Lobsang Dhargyal, \textsl{arXiv:1711.09772 and references there in.}

\bibitem{3} Lobsang Dhargyal and S. K. Rai, \textsl{arXiv:1806.01178 and references there in.}

\bibitem{4} Lobsang Dhargyal, \textsl{arXiv:1808.06499 and references there in.}

\bibitem{5} Lobsang Dhargyal, \textsl{arXiv:1810.10611 and references there in.}



\end{thebibliography}
\end{document}